\documentclass{iopart}
\newtheorem{definition}{Definition}[section]

\begin{document}

\title{Averaging Spacetime: Where do we go from here?}

\author{R. J. van den Hoogen$^*$}
\address{Department of Mathematics, Statistics, and Computer Science, St. Francis Xavier University\\
 Antigonish, Nova Scotia, B2G 2W5, Canada\\
$^*$E-mail: rvandenh@stfx.ca\\
www.stfx.ca}

\begin{abstract}
   The construction of an averaged theory of gravity based on Einstein's General Relativity is
  very difficult due to the non-linear nature of the gravitational field equations. This problem
  is further exacerbated by the difficulty in defining a mathematically precise covariant averaging
  procedure for tensor fields over differentiable manifolds.  Together, these two ideas have been
  called the averaging problem for General Relativity.  In the first part of the talk, an attempt
  to review some the various approaches to this problem will be given, highlighting strengths, weaknesses,
  and commonalities between them. In the second part of the talk, an argument will be made, that if
  one wishes to develop a well-defined averaging procedure, one may choose to parallel transport along
  geodesics with respect to the Levi-Cevita connection or, use the Weitzenb\"ock connection and
  ensure the transportation is independent of path. The talk concludes with some open questions to
  generate further discussion.

\end{abstract}




\section{Smoothing of Space-time: What are the problems?}

The current widely accepted, standard cosmological model is a Universe that is homogeneous and isotropic on the largest of length scales filled with a mixture of baryonic matter, electromagnetic radiation, neutrinos, dark matter and dark energy. The detailed measurements of the cosmic microwave background via the WMAP program and other observational programs lend credence to this model. Interestingly enough, the two dark components are estimated to make up approximately 95\% of the matter content of the Universe today and yet we have no true understanding of their composition or nature. Actually, the dark components can only be observed through their gravitational effects on baryonic matter and photons and are not directly observed. Can there be an alternative description for these observational effects that does not assume the existence of these mysterious dark quantities?

One possibility is that both dark matter and dark energy are artefacts of some effective averaged theory of gravitation. If one employs Einstein's theory of General Relativity as our theory of gravity, then the resulting cosmological model based on the assumptions of homogeneity and isotropy is described by a single function of time.  This idealized model is mathematically elegant and consistent with the observations on cosmological scales (provided one adds sufficient quantities of dark matter and dark energy to the model). However, can one simply ignore the structure and inhomogeneity that is present on smaller scales? Voids, walls, filaments, clusters, and super clusters are all examples of such structures that can be observed. Indeed, the smaller the scale, the larger the inhomogeneity.   Can one neglect the effects of these inhomogeneities on our smoothed out idealized model? If not, how does one incorporate these effects into the model? Can the dynamical effects of these inhomogeneities be used as alternative explanation to dark matter and/or dark energy that is required in the standard model? Is General Relativity the correct theory of gravity to use on cosmological scales?

The Einstein Field Equations (EFEs) of General Relativity
$$G_{\alpha\beta}(g)=\kappa T_{\alpha\beta}$$
are currently used to describe the dynamics in both the realm of cosmology and the realm of the solar or stellar systems.  However, most tests of General Relativity take place only on the scale of the solar system or stellar systems where Einstein's theory of General Relativity is consistent with all tests done to date\cite{Will}.  Unfortunately, on much larger length scales, General Relativity appears to have some problems.  Current gravitational and cosmological models using General Relativity fail to match observations on both the galactic scale (rotation curves of galaxies) and cosmological scales (acceleration of the universe) without the addition of the ``dark'' entities, matter and energy.

What then might be the cause of the problem?  What is different about the usual assumptions put into each model, solar system and cosmology.  For solar system models, the right hand side is assumed to be an idealized, centralized, isolated source, however, for cosmological models, the right hand side is commonly modeled as a fluid.  This means an averaging or smoothing procedure has been employed without a corresponding averaging or smoothing procedure on the left hand side of the equations.  Shirokov and Fisher \cite{shirokovfisher} had recognized the problem in the early 1960's, however it was not a well known issue until much later.  It was only when George Ellis \cite{Ellis} gave a detailed description of the issues did the averaging problem receive the attention necessary to become a fundamental problem in mathematical cosmology.  Both Ellis \cite{Ellis} and Shirokov and Fisher\cite{shirokovfisher}, suggested modified gravitational field equations for cosmology of the form
$$\overline G_{\alpha\beta}[g]=\kappa\overline T_{\alpha\beta}=\kappa T^{\rm fluid}_{\alpha\beta}$$ where an overbar indicates that an averaging or smoothing process has taken place. However, this suggestion requires a response to the problem of averaging tensor fields on a manifold.  {\bf Problem A}: How does one {\bf A}verage tensor fields on a manifold?  This is a non-trivial challenge due to the fact that there is no straight forward procedure for adding and comparing tensors located at different points on a differentiable manifold.

Assuming that one has addressed problem {\bf A}, one is left with an additional dilemma. Is it possible to relate $\overline G_{\alpha\beta}[g]$ with $G_{\alpha\beta}[\overline g]$?  Can we simply assume $\overline G_{\alpha\beta}[g]=G_{\alpha\beta}[\overline g]$?  In general, the answer is no, due to non-linearity of the EFEs. Both Shirokov and Fisher\cite{shirokovfisher} and Ellis\cite{Ellis} introduce a Gravitational Correlation Tensor, $C_{\alpha\beta}=\overline G_{\alpha\beta}[g]-G_{\alpha\beta}[\overline g]$ so that the averaged field equations become
$$G_{\alpha\beta}[\overline g]+C_{\alpha\beta}=\kappa T^{\rm fluid}_{\alpha\beta}$$ but this leads us to one additional problem.  {\bf Problem C}: What is the nature of the gravitational {\bf C}orrelation , $C_{\alpha\beta}$?

Here we briefly review some of the approaches that have attempted to address problems {\bf A} and {\bf C}.


\section{Proposed solutions to problems A and C}

\subsection{Space-time Averaging}
In the 1960's, Shirokov and Fisher\cite{shirokovfisher} proposed a space-time averaging procedure of the form
$$\overline T^{\alpha}_{\beta}(x)=\frac{\int_{\xi\in\Sigma_x}T^{\alpha}_{\beta}(x+\xi)\sqrt{-g(x+\xi)}\,d^4\xi}{\int_{\xi\in\Sigma_x}\sqrt{-g(x+\xi)}\,d^4\xi}$$
where $x$ is the location of the macro-observer (center of averaging region) and $\xi$ is the location of the micro-observer with respect to $x$.
Unfortunately, the proposed averaging procedure is {\bf non-covariant}, that is, the proposed average of a tensor is no longer a tensor. Further, in their analysis, the nature of the gravitational correlation, $C_{\alpha\beta}$, was determined through perturbative techniques and therefore the gravitational correlation is {\bf non-exact}. Consequently, the observations and results in this paper must be interpreted with care and consideration of these two weaknesses.

The first covariant averaging or smoothing procedure appears to be found in a series of papers by Isaacson \cite{Isaacson} in the late 1960's.  Isaacson realized the problem of trying to add tensors at different points and therefore proposed an averaging procedure of the form
$$\overline T^{\alpha}_{\beta}(x)=\int_{all \ space}g_{\alpha}^{\alpha'}(x,x')g^{\beta}_{\beta'}(x,x')T^{\alpha'}_{\beta'}(x')f(x,x')\,d^4x'$$
where $f(x,x')$ is a weighting function that falls to zero when $x$ and $x'$ differ more than some selected averaging scale and
$\int_{all \ space}f(x,x')\,d^4x'=1$. The quantity $g_{\alpha}^{\alpha'}(x,x')$ is the parallel propagator along the geodesic joining $x$ and $x'$ \cite{Synge1966,Poisson2004}. The primary purpose of these papers was not explicitly for cosmology but to calculate the effective gravitational radiation in the high-frequency limit. Although the averaging procedure is covariant, the gravitational correlation was determined perturbatively and therefore the result is {\bf non-exact}.  It must be noted that in such an averaging scheme, the average of the metric is once again the same quantity, i.e, $\overline{g}_{\alpha\beta} = g_{\alpha\beta}$.  This observation leads us to question what geometrical object should be averaged to determine the effective averaged space-time.

There have been a number of other space-time averaging procedures proposed since the 1980's.  For example, Noonan \cite{Noonan} introduced the concept of micro and macro observers and the idea of duality. Unfortunately Noonan employed a non-covariant space-time averaging procedure and perturbatively determined the gravitational correlation. Therefore his results are both {\bf non-covariant} and {\bf non-exact}. Stoeger, Helmi, and Torres \cite{Stoeger} employed a space-time averaging procedure which was equivalent to that of Shirokov and Fisher, hence the calculations are {\bf non-covariant}.  Since they applied the averaging procedure to a linearized model of gravity and a perturbed Friedman-Robertson-Walker model the results found therein are {\bf non-exact}.

In the early 1990's, Zalaletdinov \cite{Zalaletdinov} developed an averaging procedure that is similar to that proposed by Isaacson.  To achieve a covariant averaging procedure, Zalaletdinov assumed the existence of bi-local transport operators ${\mathcal A}_{\alpha}^{\alpha'}(x,x')$ that map a tensor at $x'$ to a tensor at a point $x$ in a manner that resembles the geodesic parallel transporter used by Isaacson. With some assumptions on the properties of the bi-local transport operators ${\mathcal A}_{\alpha}^{\alpha'}(x,x')$, Zalaletdinov defined a covariant space-time averaging procedure as
$$\overline T_{\alpha}^{\beta}(x)=\frac{\int_{x'\in\Sigma_x}{\mathcal A}_{\alpha}^{\alpha'}(x,x'){\mathcal A}^{\beta}_{\beta'}(x,x')T^{\alpha'}_{\beta'}(x')\sqrt{-g(x')}\,d^4x'}{\int_{x'\in\Sigma_x}\sqrt{-g(x')}\,d^4x'}$$
Instead of immediately averaging the EFEs, Zalaletdinov applied the averaging procedure to the Cartan structure equations.
To determine the nature of the gravitational correlation, Zalaletdinov first defined a second order connection correlation tensor.
With a few more assumptions, he was able to obtain splitting rules for various products of the Riemann tensor and the metric as well as a set of equations to determine parts of the connection correlation tensor.  With these results, Zalaletdinov was able to apply the averaging procedure to EFE's to calculate the gravitational correlation in a non-perturbative manner.  The resulting theory of Macroscopic Gravity is at present the only proposal in which the gravitational correlation is determined exactly and is based on a covariant averaging procedure.  Particular cosmological models built on the theory of Macroscopic Gravity suggest that the gravitational correlation behaves as a spatial curvature term\cite{CPZ,vandenHoogen2009}.  However, there are some weaknesses in the approach of Zalaletdinov.  One of the weaknesses is the assumed existence and ill understood purpose of the bi-local transport operators. A second weakness results from the fact that the connection correlation tensor is not completely determined, and therefore there is still some ambiguity in the nature of the gravitational correlation \cite{vandenHoogen2009}. The plethora of assumptions made to determine the gravitational correlation as found in the development of Macroscopic Gravity is also some cause for concern.

\subsection{Space Averaging}

Futamase \cite{Futamase1} first studied the effects of averaging in a cosmological setting by employing a perturbative expansion of the metric and a simple spatial averaging procedure to determine the gravitational correlation. Futamase's preliminary analysis is consequently {\bf non-covariant} and {\bf non-exact}.  In a later paper, Futamase\cite{Futamase2} realizing the weakness in his previous work improved upon the averaging scheme by using the parallel propagator along a geodesic on 3-surface in a similar fashion to the work of Isaacson\cite{Isaacson}.  While the procedure is not strictly covariant, the averaging of the 3-geometry in this way is a covariant averaging process.  The gravitational correlation was determined perturbatively, and hence the final result in this later work is also {\bf non-exact}.  However, even given the shortcomings in the overall analysis, Futamase determined that the dominant dynamical behavior of the gravitational correlation term behaves like a negative spatial curvature term in agreement with the conclusions found in \cite{CPZ,vandenHoogen2009} based upon the covariant and exact procedure of Zalaletdinov\cite{Zalaletdinov}.

Kasai \cite{Kasai93} approached the problem in a slightly different manner to that of Futamase\cite{Futamase1,Futamase2}. Kasai foliated the space-time by flow orthogonal hyper-surfaces having 3-metric $g_{ij}$ and concentrated his efforts on averaging a particular scalar quantity on the 3-dimensional hyper-surface.  Kasai assumed an inhomogeneous and irrotational dust model with energy density $\rho$, and calculated the spatial average of the energy density which then defined the background energy density, $\rho_b$ of the averaged space-time.  With the assumption that the inhomogeneous dust model is an FRW dust model on average, Kasai was now able to define the scale factor $a(t)$ of the background geometry via the relation $8\pi G\rho_b=3 a^{-3}$.   With the background geometrical quantities now known functions of the background energy density, Kasai defined two geometrical correlations [extrinsic curvature correlation and the Ricci 3-Curvature correlation], and a density contrast function which completely characterize the gravitational correlation.  Kasai determined the necessary conditions for the EFE's of inhomogeneous and irrotational dust models to have the form of a dust FRW on average. If these conditions were met, then some part of the gravitational correlation was determined. One of the strengths of the approach of Kasai is the fact that it is fully non-linear and exact.  Unfortunately, the procedure applies only to irrotational dust models that are FRW on average. The weakness with this assumption is that the background geometry is fixed a priori and not determined via the averaging process.  In addition, only part of gravitational correlation is explicitly calculated, some part is left undetermined.

Buchert \cite{Buchert1,Buchert2} improves upon the analysis of Kasai \cite{Kasai93} by not fixing explicitly the background geometry a priori and concentrating his efforts on averaging scalar quantities only on the 3-dimensional hyper-surface. Buchert originally assumed that the cosmological model was an inhomogeneous dust model\cite{Buchert1} foliated by flow orthogonal hyper-surfaces with 3-metric $g_{ij}$, but later expanded the analysis to perfect fluid dust models\cite{Buchert2}.  Buchert defined the spatial averaging operation
$$\overline{T}(X^i,t)=\frac{1}{V_D}\int_{D}\, T(X^i,t) \sqrt{det(g_{ij})} d^3 X$$
which is suitable for scalars on the 3-dimensional hyper-surface. Buchert defined the volume scale factor $a_D(t)=\left(\frac{V_D(t)}{V_{D_0}}\right)^{1/3}$ and applied the averaging procedure to the scalar parts of the EFE's. The two resulting equations now contain three unknown functions. The gravitational correlation is not completely determined. This observation is the primary weakness in the Buchert approach.  Since the averaging procedure can only be applied to the scalar parts of the EFE's, the tensorial parts of the EFE's were ignored, and therefore only part of the gravitational correlation is determined in this way.  However, given the weaknesses in the approach, the analysis indicates that the gravitational correlation term behaves as a curvature term\cite{Buchert1} in some instances.

\subsection{Other Promising Approaches}

As an alternative to the approaches defined above, Boersma \cite{Boersma} defines a general averaging operator $\hat A$ and assumes that the Robertson-Walker geometry is a stable fixed point of this averaging operator.  He further defines the linearized operator, $\hat A^{(1)}$ for which stability of the averaging operation implies $ \hat A^{\infty}[\delta g_{\mu\nu}] = \lim_{n\to\infty} \hat A^{(1)n}[\delta g_{\mu\nu}]$.  He then shows that linearized averaging operation for metric perturbations, can be defined as a spatial averaging operation for scalars applied to $\delta g_{00}$ and $\delta g^i_i$ in synchronous coordinates. Unfortunately, the gravitational correlation is not completely determined.

Employing the space-time averaging scheme of Zalaletdinov\cite{Zalaletdinov} and using a special limiting argument, Paranjape and Singh \cite{Paranjape} successfully constructed a spatial averaging procedure that is based on a covariant and exact averaging procedure. Restricting their analysis to a cosmology in carefully selected gauges they observed that their equations agreed with, but were not the same as, the equations found in Buchert's spatial averaging scheme for scalars described above.

More recently, Gasperini, Marozzi, and Veneziano \cite{Gasperini} proposed a Gauge invariant definition for averaging tensor fields.  They introduce a {\em Window Function} similar in nature to Isaacsons weighting function $f(x,x')$.  They make the argument that a 3 dimensional spatial averaging can be determined from the 4 dimensional averaging scheme with an appropriate choice of the {\em Window Function}. Unfortunately, they did not apply the averaging procedure to EFE's and therefore have not attempted to address the nature of the gravitational correlation or problem {\bf C}.

In 2004, Debbasch \cite{Debbasch} made arguments that an ensemble averaging procedure is necessary to average geometries, Unfortunately, no attempt to explicitly calculate the gravitational correlation resulting from averaging the EFEs is made. More recently, Sussman \cite{Sussman} defined quasi-local scalars in a class of inhomogeneous cosmological models, and was successful in averaging the scalar parts of the  EFEs: in this way, it was similar in spirit to approach proposed by Buchert. Behrend \cite{Behrend}, has proposed a space-time averaging procedure based on the idea of a maximally smooth tetrad field which determines the averaged metric. Hehl and Mashoon \cite{hehl}, developed a non-local theory of gravity based on a causal space-time averaged theory of gravity using GR${}_{||}$ which also shows some considerable promise. Khosravi, Mansouri and Kourkchi \cite{Khosravi} have constructed some preliminary but important contributions to the ideas of ``on'' and ``in'' light cone averaging. Indeed, this idea has been further developed by Coley \cite{Coley09} in which he discusses the need for {\em Light-cone Averaging} and averages the Raychaudhuri equation on the null cone in the same spirit of the 1+3 approach of Buchert.


\section{Another proposal to solve Problem A}

\subsection{The Parallel Transport Problem}

Averaging involves integration/summation of tensor fields located at finitely separated points. This process is not straightforward on an arbitrary affinely connected and curved manifold. Specifically, one requires a notion of parallel transport of a tensor at a point $x'$ along some curve $C_{x'\to x}$ to a base point $x$ to which we assign the average value in a unique and well defined manner.  In general, the value of the tensor at $x'$ when parallel transported to $x$ is dependent upon the path $C_{x'\to x}$, it takes to reach $x$, and the affine connection used to facilitate the parallel transport.

Two possible choices emerge that yield well defined transportation processes.    The first choice is to parallel transport tensor fields along well defined curves with respect to a chosen connection, for example, parallel transport along geodesics with respect to the Levi-Cevita connection.  Alternatively, and perhaps slightly more natural, one could select a connection such that the transportation process is independent of the path.  Here we briefly investigate the two possible choices.

Any tensor density field $T$ with components $T^{\alpha_1\dots\alpha_n}_{\beta_1\dots\beta_m}(x)$ can be parallel transported along a curve $C_{x'\to x}$ parameterized by $z^\gamma(\tau)$ from a point $x'=z^\gamma(\tau_0)$ to a point $x=z^\gamma(\tau)$ via the {\bf elementary parallel propagators} $P_w$, $P^{\alpha}_{\alpha'}$ and $P_{\alpha}^{\alpha'}$
    $$T^{\alpha_1\dots\alpha_n}_{\beta_1\dots\beta_m}(x)= P_wP^{\alpha_1}_{\alpha'_1}\dots P^{\alpha_n}_{\alpha_n'}P_{\beta_1}^{\beta'_1}\dots P_{\beta_m}^{\beta'_m} T^{\alpha'_1\dots\alpha'_n}_{\beta'_1\dots\beta'_m}(x')$$
There is an excellent introduction to these elementary parallel propagators in the text by Pleba\'nski and Krasi\'nski \cite{Plebanski}.  Some of the properties of the {\bf elementary parallel propagators} are described briefly below.
\begin{itemize}
\item The {\bf elementary parallel propagators} satisfy the differential equations $$ \frac{D}{D\tau}\left[P_w(\tau)\right]=0,\quad \frac{D}{D\tau}\left[P^{\alpha}_{\alpha'}(\tau)\right]=0, \quad \frac{D}{D\tau}\left[P_{\alpha}^{\alpha'}(\tau)\right] = 0$$
\item The {\bf elementary parallel propagators} have the following initial values:  $$P_w(0)=1, \quad P^{\alpha}_{\alpha'}(0) = \delta^\alpha_{\alpha'}, \quad P_{\alpha}^{\alpha'}(0) = \delta^{\alpha'}_{\alpha}$$
\item The {\bf elementary parallel propagators} $P_w$, $P^{\alpha}_{\alpha'}$ and $P_{\alpha}^{\alpha'}$ are functions of the connection, the endpoints $x'$ and $x$ and the curve $C$ parameterized by $z^\gamma(\tau)$ connecting the two points.
\item Inverse Property: $P^{\alpha}_{\alpha'}P_{\alpha}^{\beta'}=\delta^{\beta'}_{\alpha'}$ and $P^{\alpha}_{\alpha'}P_{\beta}^{\alpha'}=\delta^{\alpha}_{\beta}$.
\end{itemize}

\subsection{Choice 1: Parallel Transport along Geodesic}

To begin, we must first select a unique curve that connects the points $x$ and $x'$ and a connection: for our purposes, we choose the geodesic and the Levi-Cevita connection.  The geodesic is a ``natural'' choice as there are no other ``natural'' curves that connect $x'$ and $x$. In Riemannian space, the geodesic is the shortest and straightest path connecting points $x'$ and $x$. A weakness in this approach is the assumption that a unique geodesic exists connecting $x'$ and $x$.  The {\bf elementary parallel propagators} no longer depend on an arbitrary curve and may be considered as functions of the endpoints $x'$ and $x$.  These special parallel propagators are usually denoted with a lower case $g$, i.e, $g_w(x,x'),g^{\alpha}_{\alpha'}(x,x'),g_{\alpha}^{\alpha'}(x,x')$ to conform with standard notation for such objects\cite{Synge1966,Poisson2004}.

\subsection{Choice 2: Path Independent Parallel Transport}

It is known that parallel transport is independent of path if and only if the curvature of the connection is zero.  Therefore if one desires a path independent parallel transport, then one cannot use the Levi-Cevita connection to facilitate the parallel transport unless one is in flat space.  One must employ a different connection, one in particular that has zero curvature.  It is also known, that if the curvature is zero, then there exists $n$ covariantly constant vector fields.  Let $e_i{}^\alpha$ $(i=1,\dots, n)$ be $n$ linearly independent vector fields and assume they are covariantly constant with respect to some, as yet, undetermined connection. This requirement uniquely defines the affine connection $W^{\alpha}{}_{\beta\gamma}=e_{i}{}^{\alpha}e^i{}_{\beta,\gamma}$ better known as a Weitzenb\"ock connection.  In this case, the {\bf elementary parallel propagators} are factorable and can be shown to have the form
\begin{eqnarray*}
 P_w(x,x')&=&\left(\frac{e(x)}{e(x')}\right)^{w}\qquad e=det(e_i{}^{\alpha})\\
 P^{\alpha}_{\alpha'}(x,x')&=& e_i{}^{\alpha}(x)e^i{}_{\alpha'}(x') \\
 P_{\alpha}^{\alpha'}(x,x')&=& e^i{}_{\alpha}(x)e_i{}^{\alpha'}(x')
\end{eqnarray*}

\subsection{Covariant Averaging Procedure}

Two options are proposed to define a covariant averaging procedure.
The first is to parallel transport along the geodesic with respect to Levi-Cevita connection, in which case the curve and connection are now uniquely chosen.   This was the approach used by Isaacson\cite{Isaacson} in his investigations.  The difficulty is that one is required to determine the geodesic first, and then explicitly determine the parallel transporter via the initial value problem for the vector field to determine the nature of the parallel transporter. The second choice, is to choose an affine connection that yields a {\em Path Independent} transportation process.
Here, one parallel transports with respect to the Weitzenb\"ock connection in which case the parallel transporter has a nice form and can be calculated quite easily.  This approach closely resembles the approach used by Zalaletdinov in his investigations into averaging space-time \cite{Zalaletdinov}.
Nonetheless, in either case, one can now integrate vector and/or tensor fields over compact regions of the manifold, and consequently, one can define an averaging or smoothing procedure.

\begin{definition}[Averaging/Smoothing Procedure]
Let $\mathcal M$ be a simply connected 4-dimensional manifold having a metric with Lorentzian signature.
Let $T^{\alpha}_\beta(x)$ be a continuous tensor field defined on some simply connected region ${\mathcal R}\subset {\mathcal M}$.  Let $\Sigma_x$ be a compact subset of $\mathcal R$ at supporting point $x$.  We define the average of the tensor field $T{}^\alpha_{\beta}(x)$, denoted ${\overline T}^\alpha_{\beta}(x)$, as the definite integral at supporting point $x$,
$${\overline T}{}^\alpha_{\beta}(x)\equiv\frac{1}{V_{\Sigma_x}}\int_{x'\in\Sigma_x}\, P^\alpha_{\alpha'}(x,x')P^{\beta'}_{\beta}(x,x')T^{\alpha'}_{\beta'}(x') \sqrt{-g(x')}d^4x'$$
where $$V_{\Sigma_x}=\int_{x'\in\Sigma_x}\sqrt{-g(x')}d^4x'$$
Note: one can substitute the geodesic transporter $g^\alpha_{\alpha'}(x,x')$ if one decided to use geodesics and the Levi-Cevita connection. \end{definition}

If one chooses to parallel transport with respect to the Levi-Cevita connection, then since the connection is metric compatible, we observe that $\overline{g_{\alpha\beta}}=g_{\alpha\beta}$.  That is the metric is an invariant under the averaging procedure.  This observation may go against the natural idea that it is the metric that needs to be averaged. However, a rather nice observation results.  The average of the Riemann curvature tensor corresponding to a space of constant curvature is also invariant under this averaging procedure. This implies that the average of a space-time of constant curvature, is itself, i.e., spaces of constant curvature are fixed points of the averaging/smoothing process, which is indeed a natural expectation of any averaging/smoothing procedure.
If it does not make sense to average the metric to obtain a description of the {\em averaged effective space-time}, what then is the appropriate geometrical object of the microscopic space-time that should be averaged to determine an {\em averaged effective space-time}.  Is it the Levi-Cevita connection? Possibly, but highly unlikely due to potential problems with the covariance requirement. Perhaps it is the curvature tensor $R^{\alpha}_{\beta\gamma\delta}$? A much better possibility?  We have illustrated a covariant averaging procedure for tensor fields addressing problem {\bf A}.  We have {\bf not} averaged the Einstein Field Equation's of General Relativity, and therefore have not addressed problem {\bf C} of determining the gravitational correlation, so much more work to do.


\section{CONCLUDING REMARKS}

Averaging Spacetime:  Where do we go from here?  Before proceeding much further into this promising area of research, we, the research community, must address a few of the following issues.  These are not in any particular order, but are all questions that have yet to be fully explored in a satisfactory manner.
\begin{itemize}
\item On what length scale is Einstein's GR the appropriate theory of Gravitation?
\item If Einstein's GR is the appropriate theory describing the solar system and stellar systems, then what is the gravitational theory appropriate on cosmological scales?
\item Can a new theory of gravity be determined by simply averaging/smoothing the Einstein Field Equations?
\item Should the new smoothed theory appear as a totally new theory, or as GR plus new correction bits via a gravitational correlation?
\item Regarding the gravitational correlation (polarization) should it be determined
\begin{itemize}
\item through perturbative techniques, or
\item assigned via geometrical assumptions a la Zalaletdinov? or
\item left undetermined as in Buchert?
\end{itemize}
\item Should an averaging scheme define an effective averaged geometry (as in Zalaletdinov) {\bf or} should it define some averaged quantities defined on the un-averaged geometry (as in Buchert)?
\item Given that averaging/smoothing is much simpler to do in those spaces with tele-parallelism, should we average the GR$\|$ equivalent to GR?
\item In cosmology, should we ignore the effects of the micro geometry and concentrate solely on the macro, {\bf or} should we attempt to merge both the micro and macro?  For instance, in Maxwell's equations, the polarization terms are argued to be macroscopic in nature.
\item If the un-averaged space has a particular symmetry, then it makes some sense that any averaging procedure preserves these symmetries.  How is this to be achieved in any averaging/smoothing procedure?
\item One of our goals is to build better cosmological models.  With this goal in mind, should we use a fully covariant space-time averaging procedure, or one better suited to cosmology, e.g., a 1+3 splitting of the space-time?
\item Can the inhomogeneities in the un-averaged geometry manifest an effective acceleration in the averaged geometry?  This is the question that everyone would love to answer, but, this should come out of the model and not be a criteria that goes into the model.
\item Is the average of a bundle of null geodesics, a null geodesic?  What about Causality?
\item Cosmology is tested with observations, and observations take place down the Null Cone: Should we not be averaging down the Null Cone?
\item Even if one is able to develop an averaging/smoothing procedure, and one is able to determine the gravitational correlation via some procedure and/or assumptions, one is still left with the Fitting Problem of {\em How can one fit real-world observations to some smoothed out averaged model?}
\end{itemize}

Clearly, in order to proceed in developing a solution to the averaging problem, we must discuss and investigate a number of additional related issues concurrently.  Hence the purpose of this session at the conference.

\section*{Acknowledgments}
This research was supported by University Council on Research and the office of the Dean of Science at St. Francis Xavier University and a Discovery Grant from the Natural Sciences and Engineering Research Council of Canada.  The author would also like to thank Dr. Juliane Behrend for her comments.

\end{document}